\documentclass{phb-proc4-auth}
\usepackage{graphicx}
\usepackage{amsmath}
\usepackage{amssymb}

\begin{document}

\begin{frontmatter}

\title{Spin-lattice coupling effects in the Holstein double-exchange model}

\author[syd]{Alexander Wei{\ss}e},
\author[hgw]{Holger Fehske},
\author[lei]{Dieter Ihle}$^{,*}$,
%\ead{dieter.ihle@itp.uni-leipzig.de}
%\thanks
\corauth{Corresponding Author.\\
  phone:  +49-341-97-32433; fax: +49-341-97-32548.\\
  {\it e-mail:} dieter.ihle@itp.uni-leipzig.de} 
\address[syd]{School of Physics, The University of
  New South Wales, Sydney NSW 2052, Australia} 
\address[hgw]{Institut f\"ur Physik,
  Ernst-Moritz-Arndt Universit\"at Greifswald, 17487 Greifswald,
  Germany}
\address[lei]{Institut
  f\"ur Theoretische Physik, Universit\"at Leipzig, Augustusplatz 10 -
  11, 04109 Leipzig, Germany} 
\begin{abstract}
  Based on the Holstein double-exchange model and a highly efficient
  single cluster Monte Carlo approach we study the interplay of
  double-exchange and polaron effects in doped colossal
  magneto-resistance (CMR) manganites. The CMR transition is shown to
  be appreciably influenced by lattice polaron formation. 
\end{abstract}


\end{frontmatter}

Over the last decade the magnetic and transport properties of CMR
manganites ($\rm La_{1-x}[Ca,Sr]_xMnO_3$ with $0.2\lesssim
x\lesssim 0.5$) have attracted a considerable amount of research
activity, and in particular polaronic features near the transition
from the ferromagnetic to the paramagnetic phase remain to be an
intensely studied subject. A realistic description of the observed
$T_c$ and of the electrical resistivity data is complicated by the
requirement of incorporating strong electron-phonon  interactions
in addition to the magnetic double-exchange (DE)~\cite{MLS95}. In the
present study we aim at developing realistic effective models for the
spin lattice interaction and appropriate Monte Carlo (MC) techniques for
their simulation. Our starting point is the Holstein-DE model
%% \begin{multline}\label{holde}
%%   H = -\sum_{\langle i,j \rangle} t_{ij} c_i^{\dagger} c_j^{} 
%%   - \sqrt{\varepsilon_p\omega_0}\sum_i (b_i^\dagger +b_i^{})  
%%   (1-c_i^\dagger c_i^{})\\
%%   + \omega_0\sum_i(b_i^\dagger b_i^{}+\tfrac{1}{2})\,,
%% \end{multline}
%  \begin{multline}\label{holde}
%    H = -\sum_{\langle i,j \rangle} t_{ij} c_i^{\dagger} c_j^{} 
%    - \sqrt{\varepsilon_p\omega_0}\sum_i (b_i^\dagger +b_i^{}) 
%    c_i^\dagger c_i^{}\\
%    + \omega_0\sum_i b_i^\dagger b_i^{}\,,
%  \end{multline}
\begin{equation}\label{holde}
  H = -\sum_{\langle i,j \rangle} t_{ij} c_i^{\dagger} c_j^{} 
  - \sqrt{\varepsilon_p\omega_0}\sum_i (b_i^\dagger +b_i^{}) 
  c_i^\dagger c_i^{}
  + \omega_0\sum_i b_i^\dagger b_i^{}\,,
\end{equation}
where the first term describes the well known DE 
interaction, characterised by the transfer amplitude $t_{ij}^{} =
\cos\tfrac{\theta_i - \theta_j}{2} \cos\tfrac{\phi_i - \phi_j}{2} +
i\cos\tfrac{\theta_i + \theta_j}{2} \sin\tfrac{\phi_i - \phi_j}{2}$
which depends on the classical spin variables $\{\theta_i,\phi_i\}$.
The second term accounts for a local coupling ($\propto
\varepsilon_p$) of doped carriers to a dispersionless optical phonon mode
with frequency $\omega_0$, and the last term refers to the dynamics of
the harmonic lattice. 

As a first step let us focus on the numerical solution of the
DE part, which is characterised by non-interacting
fermions coupled to classical spin degrees of freedom. In a 
MC simulation of such types of models the calculation of the
fermionic energy contribution, which depends on the classical 
degrees of freedom, is
usually the most time consuming part, and an efficient MC algorithm
should therefore evaluate the fermionic trace as fast and as seldom as
possible. The first requirement can be matched by using Chebyshev
expansion and kernel polynomial methods~\cite{SRVK96}, but
so far this approach was combined only with standard Metropolis
single-spin updates~\cite{MF99*MF00}. For the second requirement a 
hybrid MC involving classical time evolution of an effective
spin model and an approximate diagonalisation of the fermionic
problem was suggested~\cite{AFGLM01}.  However, both approaches have a
few drawbacks, the first suffers from the frequent evaluation of the
fermionic trace, the latter is rather complicated since it involves a
molecular dynamics type simulation of the classical degrees of freedom.

\begin{figure}
  \begin{center}
    \includegraphics[width=0.9\linewidth]{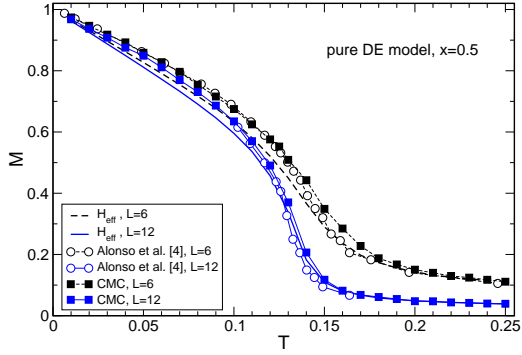}
  \end{center}
  \caption{Comparison of magnetisation data for the pure DE 
    model at $x=0.5$, obtained with the effective
    model~\eqref{heffde}, the hybrid MC approach~\cite{AFGLM01}, and
    the new CMC method. Here and in the following all energies are 
    measured in units of $t$.}\label{fig1}
\end{figure}

As an alternative we propose a combination of kernel polynomial 
expansion and cluster MC (CMC)~\cite{Ja98b*Kr03}, which is known to be the
fastest approach to non-frustrated classical models. Averaging the
first part of~\eqref{holde} over the fermionic hopping we can define
an approximate classical spin Hamiltonian
\begin{equation}\label{heffde}
  H_{\text{eff}} = - J_{\text{eff}} \sum_{\langle ij \rangle} 
  \sqrt{1 + \vec S_i\cdot \vec S_j}\,,
\end{equation}
which can be easily simulated with a rejection-free Wolff~\cite{Wo89}
single cluster type approach. Setting $J_{\text{eff}} = x (1-x)
/\sqrt{2}$, where $x$ is the hole concentration, the magnetisation
data and critical temperatures of this approximate model agree
surprisingly well with the full DE system (cf. Fig.~\ref{fig1}). Thus,
the latter can efficiently be simulated by the new hybrid approach:
(i) all spins of the system are altered by multiple cluster flips,
where preferably $J_{\text{eff}}$ is random with $\langle
J_{\text{eff}}\rangle = x(1-x)/\sqrt{2}$), (ii) the fermionic energy
is evaluated using the kernel polynomial method, and (iii) the new
spin configuration is accepted with probability $\tilde P(a\to b) =
\min\left(1, \frac{W(b) A(b\to a)}{W(a) A(a\to b)}\right)$. Here
$W(a)$, $W(b)$ are the Boltzmann weights of the spin configurations
$a$, $b$, and $A(a\to b)$, $A(b\to a)$ are the accumulated {\em a
  priori} probabilities of the cluster flips connecting $a$ and
$b$~\cite{Ja98b*Kr03}. Of course, the MC update is no longer
rejection-free, but since $H_{\text{eff}}$ yields very good estimates
for $H$, $\tilde P(a\to b)$ is sufficiently high for a reliable and
fast simulation of reasonably large clusters. This allows for a
confirmation and possible future extension of previous results for the
DE model~\cite{AFGLM01}.

\begin{figure}
  \begin{center}
    \includegraphics[width=\linewidth]{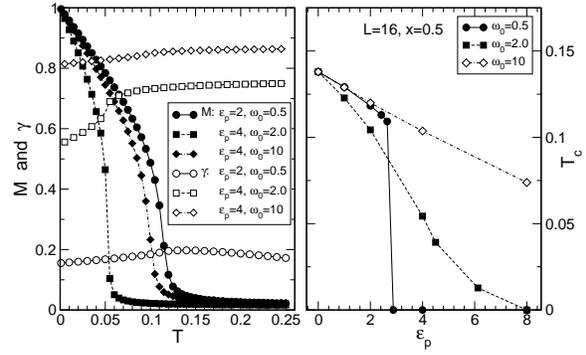}
  \end{center}
  \caption{Left: Magnetisation data (filled symbols) 
    and polaron parameter $\gamma$ (open symbols) for 
    typical values of $\varepsilon_p$ and $\omega_0$. Right:
    Dependence of the critical temperature on the  electron-phonon interaction
    strength.}\label{fig2}
\end{figure}

Encouraged by the above findings we can now rely entirely on the
effective spin model $H_{\text{eff}}$ and include lattice polaron
effects on a variational level. Applying the modified variational
Lang-Firsov transformation~\cite{FRWM95}, from~\eqref{holde}
and~\eqref{heffde} we obtain the effective polaron-DE model,
\begin{multline}\label{hampol}
  H_p = - J_{\text{eff}}\, e^{-\tfrac{\gamma^2\varepsilon_p}{\omega_0}
  \coth\big[\tfrac{\beta\omega_0}{2}\big]}  \sum_{\langle ij \rangle} 
  \sqrt{1 + \vec S_i\cdot \vec S_j}\\
%\ H_{\text{eff}}\\
  + \tfrac{N \omega_0}{e^{\beta \omega_0} -1}
  + N \varepsilon_p x \big[(1-\gamma)^2 (1-x) - 1\big]\,,
\end{multline}
where $\gamma$ is a variational parameter which measures the
importance of the polaron effect ($0\leq\gamma\leq 1$). 
The above model is treated by single
cluster MC, where the optimal $\gamma$ is adjusted after each cluster
flip. Note that the spin-lattice part of~\eqref{hampol} is symmetric
with respect to $x=0.5$, i.e., to explain the asymmetry of the
manganite phase diagram additional Jahn-Teller type lattice
interactions need to be included. Figure~\ref{fig2} shows the
magnetisation as well as $\gamma$ for a few typical parameter sets.
Clearly the polaronic effect (larger $\gamma$) is most pronounced near
and above the critical temperature. As expected, the electron-phonon 
interaction reduces the critical temperature, and for small phonon frequencies 
(adiabatic case) the ferromagnetic phase may cease to exist
completely. Potential future extensions of the present work could
include Jahn-Teller modes~\cite{WF02} and studies of the
conductivity, e.g., along the lines of Ref.~\cite{Ih85}. In addition,
the formation of magnetic polarons in the DE model should be 
accessible within the proposed CMC approach.


\begin{thebibliography}{10}

\bibitem{MLS95}
A. J. Millis, P. B. Littlewood, and B. I. Shraiman,
Phys. Rev. Lett. {\bf 74}, 5144 (1995).

\bibitem{SRVK96}
%R.~N. Silver, H.~R\"oder,  Int. J. Mod. Phys. C {\bf 5}, 935 (1994);
R.~N. Silver, H.~R\"oder, A.~F. Voter, D.~J. Kress,  J. of Comp.
  Phys. {\bf 124}, 115 (1996).
  
\bibitem{MF99*MF00} 
Y.~Motome, N.~Furukawa, J. Phys. Soc. Jpn. {\bf 68}, 3853 (1999); 
%Y.~Motome, N.~Furukawa, J. Phys. Soc. Jpn. 
{\it ibid.} {\bf 69}, 3785 (2000).

\bibitem{AFGLM01} J.~L. Alonso, L.~A. Fern{\'a}ndez, F.~Guinea,
  V.~Laliena, V.~Mart{\'i}n-Mayor, Nucl. Phys. B 596 (2001) 587.

\bibitem{Ja98b*Kr03}
W.~Janke, Math. and Comput. in Simul. {\bf 47},  329 (1998); 
W.~Krauth, cond-mat/0311623.

%% \bibitem{Kr03}
%% W.~Krauth, cond-mat/0311623.

\bibitem{Wo89}
U.~Wolff, Phys. Rev. Lett. {\bf 62}, 361
  (1989).

\bibitem{FRWM95} 
H. Fehske, H. R\"oder, G. Wellein, and A. Mistiotis, Phys. Rev. B 
{\bf 51}, 16582 (1995).

\bibitem{WF02}
A. Wei{\ss}e and H. Fehske, Eur. Phys. Jour. B 30, 487 (2002).

\bibitem{Ih85}
D. Ihle, Z. Phys. B {\bf 58}, 91 (1985).
\end{thebibliography}
\end{document}